\documentstyle[seceq,preprint,mbf,mssymb]{ptptex}  

\def\ifundefined#1{\expandafter\ifx\csname#1\endcsname\relax}
\def\ifpreprint{\ifundefined{abstsize}}

\def\bfit{\bfseries\itshape}

\ifpreprint 
 \font\titlefont=cmbx10 scaled\magstep3
 \font\instfont=cmti10 scaled\magstep1
\else
 \ifundefined{publishedin} 
   \font\titlefont=cmbx10 scaled\magstep1
 \else                                  
   \def\titlefont{}
 \fi
 \def\instfont{\small\it}
 \if 
\fi

\ifpreprint   

\def\footnote{}
\newcounter{footn}[page]
\def\fnsymbn#1{\ifcase#1\or *\or **\or ***\or ****\or *****
               \else \fi\relax}
\def\fnsymb{\fnsymbn{\value{footn}}}
\def\foot#1{\addtocounter{footn}{1}$^{\fnsymb)}$\footnotetext{
\hspace*{-20pt}\vbox{\hsize15pt\noindent\hfill$^{\fnsymb)}$} #1}}
\def\thank#1{\addtocounter{footn}{1}$^{\raise-3pt\hbox{\small\fnsymb})}$
             \footnotetext{\hspace*{-20pt}%
                         \vbox{\hsize14pt\noindent\hfill$^{\fnsymb}$}\ \ #1}}
\else
\def\foot{\footnote}
\def\thank{\footnote}
\fi

\def\eqno(#1,#2){(\ref{eq:#1,#2})}

\def\vphi{\varphi}
\def\lag{{\cal L}}
\def\lagI{\lag_{\hbox{\tiny I}}}
\def\calT{{\cal T}}
\def\tcalT{\skew3\widetilde{\calT}}
\def\hcalT{\skew3\widetilde{\tcalT}}
\def\sqr#1#2{{\vcenter{\vbox{\hrule height.#2pt
                 \hbox{\vrule width.#2pt height#1pt \kern#1pt
                       \vrule width.#2pt}
                 \hrule height.#2pt}}}}

\def\tg{\,\tilde{\! g}}  
\def\hg{\hat{g}{}}
\def\tb{\,\tilde{\! b}{}}
\def\barc{{\skew3\bar c}\hspace*{1pt}}
\def\gdel{{\mbf\delta_*}\,}
\def\Dp{D^{\hbox{\tiny (+)}}}

\def\Df{D_{\hbox{\tiny F}}}
\def\wightman#1{\langle\,#1\,\rangle}

\def\taustar#1{\wightman{\hbox{T}^*\,#1}}

\def\abemail{abe@kurims.kyoto-u.ac.jp}
\def\nnmail{nbr-nakanishi@msn.com}

\notypesetlogo  
\preprintnumber{RIMS-1228}
{
\markboth{
M.~Abe and N.~Nakanishi
}{
Perturbative or Path-Integral Approach versus Operator-Formalism Approach}

\title{
\titlefont
Perturbative or Path-Integral Approach\\
{\bfit versus\/} Operator-Formalism Approach%
}

\author{
\ifundefined{@addenda}
Mitsuo {\sc Abe}$^{1,}$\thank{E-mail: \abemail}
and Noboru {\sc Nakanishi}$^{2,}$\thank{Professor Emeritus of
Kyoto University. E-mail: \nnmail}
\else
Mitsuo {\sc Abe}\thank{E-mail: \abemail}
and Noboru {\sc Nakanishi}\thank{Professor Emeritus of
Kyoto University. E-mail: \nnmail}
\fi
}

\inst{
\instfont
$^1$Research Institute for Mathematical Sciences, Kyoto University, 
Kyoto 606-8502, Japan
\\
$^2$12-20 Asahigaoka-cho, Hirakata 573-0026, Japan
}

\recdate{
April 5, 1999
}

\abst{
In the conformal-gauge two-dimensional quantum gravity, the solution 
obtained by the perturbative or path-integral approach is compared with
the one obtained by the operator-formalism approach.
Treatments of the anomaly problem in both approaches are different.
This difference is found to be essentially caused by the fact that the
perturbative or path-integral approach is based on the T$^*$-product
(covariantized T-product), which generally violates field equations.
Indeed, this fact induces some extra one-loop Feynman diagrams, which 
would not exist unless a nonzero contribution arose from a zero field.
Some demerits of the path-integral approach are explicitly demonstrated.
}

\begin{document}

\maketitle

\section{Introduction}
Quantum field theory is undoubtedly a very successful theory except 
for its mathematically rigorous foundation.  The basic quantities of 
quantum field theory are quantum fields, generically denoted by $\vphi(x)$, 
which are operator-valued singular functions of $N$-dimensional 
space-time $x^\mu$. In the standard formulation, their properties are 
governed by the action $S$, which is an $N$-dimensional integral of 
the Lagrangian density $\lag$ (a function of quantum fields at the 
same space-time point). Field equations and canonical (anti)commutation 
relations are derived from $S$ by the standard procedure. Equal-time 
(anti)commutation relations follow from the canonical ones.
\par
Field equations and equal-time (anti)commutation relations define the 
operator properties of quantum fields, that is, it is supposed to 
determine the operator algebra of quantum fields. Next, this operator 
algebra is represented in terms of state vectors, whose totality forms 
an infinite dimensional complex linear space equipped with (indefinite) 
inner product. This way of formulating the theory is operator-formalism 
approach. Evidently, this way of thinking is most natural as the 
formulation of quantum fields.  Nevertheless, this approach has never 
been seriously considered as the standard approach to quantum field 
theory. The main reasons for this are as follows.  First, it has been 
unknown how to find the solution in this approach.  Second, it has been 
unknown how to deal with the divergence problem which arises from the 
singular nature of quantum fields.
\par
The present-day's standard approach is covariant perturbation theory.
It is based on the action $S$, rather than field equations.  Decomposing
it artificially into its free part and its interaction part, one obtains 
all Green's functions explicitly in terms of Feynman diagrams.
Furthermore, renormalization theory tells us how to deal with 
the divergence problem: divergent contributions are absorbed into 
counter terms.
\par
Path-integral formalism gives us the generating functional, $Z$, of the 
Green's functions. If path integral is defined as the infinite multiple 
integrals with respect to the expansion coefficients of quantum fields 
in terms of their free wave functions, $Z$ reproduces the perturbative 
expressions. If path integral is defined by using an expansion in terms 
of the functions other than free wave functions, it is not clear whether 
or not $Z$ yields the Green's functions which reproduce the perturbative 
ones (if expanded), but it is usually supposed that there exists 
an appropriate path-integral measure giving the expected results.  
Although we cannot accept this assumption without 
reservation,\foot{This problem is particularly relevant in quantum
gravity because gravity's free wave functions are rather artificial 
quantities. For example, the sum over all possible manifolds having 
a particular metric signature cannot be realized by a simple infinite 
multiple integral with respect to expansion coefficients because 
a definite metric signature requires to insert a product of $\theta$ 
functions of the determinant and principal minors of $g_{\mu\nu}$ into 
the path integral.} we do not strictly distinguish perturbative 
approach and path-integral one in the present paper.
\par
By taking the logarithm of $Z$, one obtains the generating functional, 
$W$, of connected Green's functions. Moreover, the effective action 
$\Gamma$, which is the generating functional of amputated proper 
Feynman diagrams, is obtained as the functional Legendre transform 
of $W$.  The renormalization is neatly carried out in $\Gamma$ and 
therefore the anomaly problem is usually discussed also in $\Gamma$. 
It should be noted here that perturbation series is not unique because 
the decomposition of the action into its free part and its interaction 
part is generally altered by a nonlinear redefinition of fields.  
Accordingly, the effective action $\Gamma$ is a quantity which generally 
changes under the redefinition of fields.  Thus, in the path-integral 
approach, neither renormalization procedure nor the anomaly problem are 
quite independent of the choice of quantum fields.  This point becomes 
crucial when unphysical fields, whose natural definitions are not 
necessarily unique, play important roles as in quantum gravity.
\par
Now, we have recently succeeded in formulating the method of finding 
the solution in the operator-formalism approach.\cite{AN1}  
Our method is as follows. From the field equations and equal-time 
(anti)commutation relations, we explicitly construct all independent 
$N$-dimensional (anti)commutation relations, by expanding them, if 
necessary, into the power series with respect to the parameters 
involved. We then calculate independent $N$-dimensional multiple 
(anti)commutators.  The representation of the field algebra in terms 
of state vectors is constructed by giving all $n$-point Wightman 
functions ($n=1,\,2,\,\ldots\,$), i.e., vacuum expectation values of 
simple products of $n$ quantum fields, $\vphi_1(x_1)\vphi_2(x_2)\cdots
\vphi_n(x_n)$, so as to be consistent with the $(n-1)$ple 
(anti)commutators and with the energy positivity 
conditions.\foot{The Wightman function is a boundary value of 
an analytic function of the variables $x_i{}^0-x_j{}^0$ $\ (i<j)$ 
from the lower half-planes.} 
Here, in contrast with the axiomatic field theory,\cite{SW} we need the 
Wightman functions involving composite fields, where a composite field 
is a product of fields at the same space-time point.  When we set some 
of space-time points coincident in a higher-point Wightman function, 
we generally encounter divergent terms, which must be simply discarded 
in such a way that the resultant be independent of the ordering of 
the constituent fields of the composite field (``generalized normal 
product rule'').  In this procedure, we do not introduce anything like 
a counter term.  This is because a well-defined representation of the 
field algebra should be free of divergence; our standpoint is similar
to that of the Lehmann-Symanzik-Zimmermann formalism,\cite{LSZ} in which 
they developed the renormalized perturbation theory without encountering 
any unrenormalized quantities.
\par
Of course, it is extremely difficult to carry out our way of finding 
the solution in realistic models.  But, fortunately, we can explicitly 
construct the exact solutions in some two-dimensional models by our 
method.\cite{AN2,AN3,AN4,AN5,Ikeda}  
Our results are seen to be quite satisfactory, but we encounter an anomalous 
phenomenon, which we call ``field-equation anomaly'', in quantum-gravity 
models\cite{AN2,AN3,AN4,AN5} (but not in gauge-theory models\cite{Ikeda}): 
By construction, our Wightman functions are consistent with all 
two-dimensional (anti)commutators but not necessarily consistent with 
nonlinear field equations because there we encounter products of fields
at the same space-time point.
In any of the quantum-gravity models which we have exactly solved so far, 
one of field equations is {\it slightly\/} violated at the level of 
representation.\ifpreprint\setcounter{footn}{0}\fi\foot{The violation is 
{\it slight\/} in the sense that an anomaly-free equation can be obtained 
by differentiating the original field equation once or twice.}
This is the field-equation anomaly.  It is different from the 
conventional anomalies which arise in connection with particular 
symmetries. Rather, as clarified in our previous work,\cite{AN6} 
various conventional anomalies\cite{AN7} are systematically explained 
on the basis of the field-equation anomaly and their ambiguities are 
shown to be caused by the nonuniqueness of perturbation theory.  
In this sense, we regard the field-equation anomaly as a more fundamental 
concept.
\par
The purpose of the present paper is to make comparison between the 
solution obtained by the perturbative approach and the one obtained 
by the operator-formalism approach explicitly in the conformal-gauge 
two-dimensional quantum gravity. The exact solution of this model 
obtained previously\cite{AN4} can be written in terms of {\it tree 
diagrams only}.  If the same were true also in the perturbative 
solution, no anomaly could be present.  As is well known, however, 
this model has the conformal anomaly except for $D=26$, where $D$ 
denotes the number of the scalar fields which can be interpreted as 
the string coordinates.  We trace the cause of this paradox and find 
that the perturbative approach induces some one-loop Feynman diagrams,
which would not exist unless a nonzero contribution arose from a zero 
field.  The cause of this strange phenomenon is found to be the use of 
T$^*$-product (covariantized T-product) of quantum 
fields\ifpreprint\setcounter{footn}{1}\fi\foot{T$^*$-product is 
a T-product modified in such a way that%
\ifpreprint
\hbox{$\,\partial_1{}\!^0\hbox{T}^{\hskip-.5pt*}\hskip-.5pt\vphi_1(x_1)
\cdots\vphi_n(x_n)
\!=\!\hbox{T}^{\hskip-.5pt*}\hskip-.5pt\partial_1{}\!^0\vphi_1(x_1)
\cdots\vphi_n(x_n)$.} 
\else
{} $\partial_1{}^0\hbox{T}^*\vphi_1(x_1)\cdots\vphi_n(x_n)
=\hbox{T}^*\partial_1{}^0\vphi_1(x_1)\cdots\vphi_n(x_n)$.
\fi 
}
in the perturbative or path-integral approach.  More generally, in 
the present paper, we clarify that various anomalous behaviors of 
this model found in the perturbative approach are caused by the use 
of T$^*$-product.
\par
In the present paper, we compare the perturbative or path-integral
approach with the operator-formalism approach in the conformal-gauge
two-dimensional quantum gravity,\ifpreprint\break\else{} \fi whose 
Lagrangian density is given in \S2.  In \S3, \S4 and \S5, respectively, 
we discuss this model by the operator-formalism approach, by the 
perturbative approach and by the path-integral approach.  
In \S6, we criticize the so-called ``FP-ghost number current anomaly''.  
The final section is devoted to discussions.

\ifpreprint
\newpage
\fi
\section{Preliminaries}
Throughout the present paper, we consider the BRS formalism of the 
conformal-gauge two-dimensional quantum gravity, in which the conformal
degree of freedom is already eliminated.\cite{Yang}  Quantum fields are 
contravariant tensor-density gravitational field $\tg^{\mu\nu}$, FP 
ghost $c^\mu$, FP anti-ghost $\barc_{\mu\nu}$, B field $\tb_{\mu\nu}$ 
and scalar fields $\phi_M\; (M=0,\,1,\,\ldots,\,D-1)$.  
Their BRS transforms are as follows:
\begin{eqnarray}
\gdel \tg^{\mu\nu}&=& \tg^{\mu\sigma}\partial_\sigma c^\nu
                     +\tg^{\nu\sigma}\partial_\sigma c^\mu
                     -\partial_\sigma(\tg^{\mu\nu}c^\sigma), \\
\gdel c^\mu       &=&-c^\sigma\partial_\sigma c^\mu, \\
\gdel \barc_{\mu\nu} &=& i\tb_{\mu\nu}, \\
\gdel \tb_{\mu\nu} &=&0,\\
\gdel \phi_M &=& -c^\sigma\partial_\sigma\phi_M.
\end{eqnarray}
Since $\tg^{\mu\nu}$ has only two degrees of freedom because det$\tg^{\mu\nu}
=-1$ , it is parametrized as
\begin{eqnarray}
\tg^{\mu\nu}&=&(\eta^{\mu\nu}+h^{\mu\nu})(1-\det{h^{\sigma\tau}})^{-1/2},
 \label{eq:2,6}
\end{eqnarray}
where $h^{\mu\nu}$ is symmetric and traceless ($\eta_{\mu\nu}h^{\mu\nu}=0$).
Correspondingly, $\barc_{\mu\nu}$ and $\tb_{\mu\nu}$ are also symmetric
and traceless.  It is convenient to rewrite any traceless symmetric tensor
$X_{\mu\nu}$ into a vector-like quantity $X^\lambda$ by
\begin{eqnarray}
X^\lambda &=& {1\over\sqrt{2}}\xi^{\lambda\mu\nu}X_{\mu\nu}, \label{eq:2,7}
\end{eqnarray}
where $\xi^{\lambda\mu\nu}=1$ for $\lambda+\mu+\nu=$even, $=0$  otherwise.
According to \eqno(2,7), we introduce $h_\lambda$, $\barc^\lambda$ and
$\tb^\lambda$.
\par
The BRS-invariant action $S=\int d^2x\,\lag$ is given by the Lagrangian
density
\begin{eqnarray}
\lag&=&{1\over\,2\,}\tg^{\mu\nu}\partial_\mu\phi_M\cdot\partial_\nu\phi^M
       -{1\over\,2\,}\tb^\lambda h_\lambda
       -{i\over\,2\,}\barc^\lambda\gdel(h_\lambda), \label{eq:2,8}
\end{eqnarray}
where $\tg^{\mu\nu}$ is given by \eqno(2,6), that is,
\begin{eqnarray}
\lag&=&\lag_0+\lagI, \label{eq:2,9}
\end{eqnarray}
with
\begin{eqnarray}
\lag_0&=& {1\over\,2\,}\eta^{\mu\nu}\partial_\mu\phi_M\cdot\partial_\nu\phi^M
         -{1\over\,2\,}\tb^\lambda h_\lambda
         -{i\over\sqrt{2}}\xi_{\lambda\mu\nu}\barc^\lambda\partial^\mu c^\nu,
         \label{eq:2,10} \\
\lagI&=& h_\lambda
       \bigg({1\over2\sqrt{2}}\xi^{\lambda\mu\nu}\partial_\mu\phi_M\cdot
             \partial_\nu\phi^M
  \if@preprint 
       -{i\over\,2\,}\xi^{\lambda\mu\nu}\xi_{\rho\mu\sigma}\barc^\rho
                          \partial_\nu c^\sigma  
       -{i\over\,2\,}\partial_\sigma\barc^\lambda\cdot c^\sigma\bigg) 
       +O(h^2).  \label{eq:2,11}  \hspace*{20pt} 
  \else     
       \!-\!{i\over\,2\,}\xi^{\lambda\mu\nu}\xi_{\rho\mu\sigma}\barc^\rho
                          \partial_\nu c^\sigma  
       \!-\!{i\over\,2\,}\partial_\sigma\barc^\lambda\cdot c^\sigma\bigg)
     \!+\!O(h^2).  \label{eq:2,11}  \hspace*{40pt}
  \fi
\end{eqnarray}
The higher-order terms $O(h^2)$ are unnecessary to be specified because they
contribute neither to field equations nor to canonical (anti)commutation
relations.  Furthermore, they give no contribution to perturbation theory.
Thus, we may discard them.
\par
It should be noted that the action is invariant under the FP-ghost 
conjugation\foot{So far, this fact has been overlooked because the 
FP antighost was treated as a tensor.}
\begin{eqnarray}
&&c^\lambda     \ \longrightarrow \  \barc^\lambda, \nonumber \\
&&\barc^\lambda \ \longrightarrow \  c^\lambda,  \nonumber \\
&&\tb^\lambda   \ \longrightarrow \  
  \tb^\lambda - i\xi^{\lambda\mu\nu}\xi_{\rho\mu\sigma}\barc^\rho
                    \partial_\nu c^\sigma 
              - i\partial_\sigma\barc^\lambda\cdot c^\sigma
  \nonumber \\
&&\hspace*{55pt}
              + i\xi^{\lambda\mu\nu}\xi_{\rho\mu\sigma}c^\rho
                    \partial_\nu\barc^\sigma
              + i\partial_\sigma c^\lambda\cdot \barc^\sigma 
            \ + \  O(h),
\end{eqnarray}
as in the de Donder-gauge case.
\par
Analysis can be much simplified by introducing light-cone coordinates
$x^\pm=$ 
$(x^0\pm x^1)/\sqrt{2}$, because then $\xi_{\mu\nu\lambda}=0$ except
$\xi_{+++}=\xi_{---}=\sqrt{2}$.  From \eqno(2,10) and \eqno(2,11) ($O(h^2)$
is omitted), we have
\begin{eqnarray}
&&\lag_0=\partial_+\phi_M\cdot\partial_-\phi^M
         +\Big[-{1\over\,2\,}\tb^+ h_+ -i\barc^+\partial_-c^+
                + \quad (\ +\ \longleftrightarrow\ -\ )\Big], 
         \label{eq:2,13} \\
&&\lagI=h_+\Big[{1\over\,2\,}\partial_+\phi_M\cdot\partial_+\phi^M
                - i\barc^+\partial_+c^+ 
                -{i\over\,2\,}\Big(\partial_+\barc^+\cdot c^+ 
                           +\partial_-\barc^+\cdot c^- \Big) \Big] \
          \nonumber \\
&&\hspace*{50pt}
                + \quad (\ +\ \longleftrightarrow\ -\ ).
         \label{eq:2,14}
\end{eqnarray}
In subsequent sections, we start with \eqno(2,9) together with \eqno(2,13)
and \eqno(2,14). For later convenience, we introduce the following notation.
\begin{eqnarray}
\tcalT^\pm &\equiv& \partial_\pm\phi_M\cdot\partial_\pm\phi^M
                    -2i\barc^\pm\partial_\pm c^\pm
                    -i\partial_\pm\barc^\pm\cdot c^\pm, \\
\hcalT^\pm &\equiv& \tcalT^\pm - i\partial_\mp\barc^\pm\cdot c^\mp.
\end{eqnarray}
Then we have
\begin{eqnarray}
 {\partial \lagI \over \partial h_\pm} &=& {1\over\,2\,}\hcalT^\pm.
\end{eqnarray}
\par
The Noether currents of the BRS invariance and the FP-ghost number 
conservation are given by
\begin{eqnarray}
j_b{}^\mp &=& -c^\pm \partial_\pm \phi_M\cdot \partial_\pm \phi^M
              -i\barc^\pm c^\pm \partial_\pm c^\pm, 
  \label{eq:2,18} \\
j_c{}^\mp &=& -i\barc^\pm c^\pm,
\end{eqnarray}
respectively.  They are of course conserved.

\ifpreprint
\newpage
\fi
\section{Operator-formalism Approach}
For the sake of comparison, we briefly review our previous results of the
exact solution obtained by the operator-formalism approach.\cite{AN4}
\par
The field equations are as follows:
\begin{eqnarray}
&&h_\pm=0, \label{eq:3,1} \\
&&\tb^\pm=\tcalT^\pm, \label{eq:3,2} \\
&&\partial_\mp X^\pm=0 \quad \hbox{for } X^\pm=c^\pm,\ \barc^\pm,\ \tb^\pm, 
  \label{eq:3,3} \\
&&\partial_+\partial_-\phi_M=0, \label{eq:3,4}
\end{eqnarray}
where \eqno(3,3) for $X^\pm=\tb^\pm$ is derived by differentiating \eqno(3,2).
Note that $c^\pm,\ \barc^\pm,\ \tb^\pm$ and $\partial_\pm\phi_M$ are the 
functions of a single variable $x^\pm$ only.
\par
Canonical quantization is carried out by taking $\phi_M$ and $c^\pm$ only as
the canonical variables. Since they are free fields, their nonvanishing %
two-dimensional (anti)\-commutators
are easily obtained; we have
\begin{eqnarray}
&&[\partial_\pm\phi_M(x),\;\phi^N(y)]
  =-{i\over\,2\,}\delta_M{}^N \delta(x^\pm-y^\pm), \label{eq:3,5} \\
&&\{c^\pm(x),\;\barc^\pm(y)\}=-\delta(x^\pm-y^\pm).
\end{eqnarray}
The commutation relations involving $\tb^\pm$ are calculated by using
\eqno(3,2):
\begin{eqnarray}
{[} \tb^\pm(x),\;\phi_M(y) ]
 &=&-i\partial_\pm\phi_M(x)\cdot\delta(x^\pm-y^\pm), \\
{[} \tb^\pm(x),\;c^\pm(y) ]
 &=&-ic^\pm(x)\delta'(x^\pm-y^\pm)
    -2i\partial_\pm c^\pm(x)\cdot\delta(x^\pm-y^\pm), \\
{[} \tb^\pm(x),\;\barc^\pm(y) ]
 &=& 2i\barc^\pm(x)\delta'(x^\pm-y^\pm)
     +i\partial_\pm\barc^\pm(x)\cdot\delta(x^\pm-y^\pm), \label{eq:3,9} \\
{[} \tb^\pm(x),\;\tb^\pm(y) ]
 &=&i[\tb^\pm(x)+\tb^\pm(y)]\delta'(x^\pm-y^\pm). \label{eq:3,10}
\end{eqnarray}
Evidently, \eqno(3,10) is the BRS transform of \eqno(3,9).
\par
Since no new operators are encountered in the right-hand sides of 
\eqno(3,5)-\eqno(3,10), we can easily calculate all multiple 
(anti)commutators explicitly.
We then construct all truncated\foot{Truncation means to drop the 
contributions from vacuum intermediate states. The {\it truncated\/} Wightman
function corresponds to the {\it connected\/} Green's function.
In the present model, the distinction between truncated and nontruncated 
appears only for $n\geqq4$.} Wightman functions so as to be consistent with
all multiple (anti)commutators under the energy positivity condition.
\par
The 1-point functions are, in principle, completely arbitrary.  But we set 
all of them equal to zero because we should not deliberately violate any of
\eqno(3,1), FP-ghost number conservation, BRS invariance and $O(D)$ symmetry.
\par
The nonvanishing {\it truncated\/} $n$-point Wightman functions are those
which consist of $(n-2)$ $\tb^\pm$'s and of either $c^\pm$ and $\barc^\pm$ 
or two $\phi_M$'s.  Diagrammatically, they are represented by tree diagrams.
Although we have explicitly constructed all of them,\cite{AN4} we here 
quote 2-point and 3-point ones only.
\par
Nonvanishing 2-point Wightman functions are
\begin{eqnarray}
&&\wightman{\phi_M(x_1)\phi^N(x_2)}
 =\delta_M{}^N \Dp(x_1-x_2),\\
&&\wightman{c^\pm(x_1)\barc^\pm(x_2)}
  =\wightman{\barc^\pm(x_1)c^\pm(x_2)}
  =-2i\partial_\pm\Dp(x_1-x_2),
\end{eqnarray}
where\foot{$\Dp(x)$ itself is infrared divergent and therefore 
requires the introduction of infrared \linebreak[3]
cutoff.~\cite{Nakanishi}}
\begin{eqnarray}
&& \partial_\pm\Dp(x)\equiv -{1\over4\pi}\cdot{1\over x^\pm-i0}.
\end{eqnarray}
Nonvanishing 3-point ones are
\begin{eqnarray}
&&\wightman{\phi_M(x_1)\tb^\pm(x_2)\phi^N(x_3)}
=-2\delta_M{}^N \partial_\pm\Dp(x_1-x_2)\cdot\partial_\pm\Dp(x_2-x_3), 
 \hspace*{30pt} 
 \label{eq:3,14} \\
&&\wightman{c^\pm(x_1)\tb^\pm(x_2)\barc^\pm(x_3)}
=8i\partial_\pm\Dp(x_1-x_2)\cdot\partial_\pm{}^2\Dp(x_2-x_3) \nonumber\\
&&\hspace*{120pt}
-4i\partial_\pm{}^2\Dp(x_1-x_2)\cdot\partial_\pm\Dp(x_2-x_3)
 \label{eq:3,15}
\end{eqnarray}
and their permutated ones, whose expressions are obtained from the above
by changing some of $\Dp(x)$'s into $-[\Dp(x)]^*$ so as to become consistent
with the energy-positivity condition (and by changing the overall sign if
the order of $c^\pm$ and $\barc^\pm$ is reversed).
\par
Our system of Wightman functions is, of course, consistent with the field
algebra defined by \eqno(3,5)-\eqno(3,10).  It is also consistent with
the BRS invariance and the FP-ghost number conservation.  It should be noted
that we need the use of the generalized normal-product rule to check the
BRS invariance.  Our system of Wightman functions is also consistent with
all {\it linear\/} field equations \eqno(3,1), \eqno(3,3) and \eqno(3,4),
but {\it not\/} with the {\it nonlinear\/} field equation \eqno(3,2).
Indeed, by using the generalized normal-product rule, we can show that
\begin{eqnarray}
\wightman{\tb^\pm(x_1)\tb^\pm(x_2)}&=&0, \\
\wightman{\tb^\pm(x_1)\tcalT^\pm(x_2)}
&=&\wightman{\tcalT^\pm(x_1)\tcalT^\pm(x_2)} \nonumber\\
&=&2(D-26)[\partial_\pm{}^2\Dp(x_1-x_2)]^2
\end{eqnarray}
in contradiction with \eqno(3,2).  Thus the field equation \eqno(3,2),
{\it modulo\/} \eqno(3,3) for $X^\pm=\tb^\pm$, is violated at the level
of the representation in terms of state vectors.
We call this matter ``field-equation anomaly''.  This phenomenon is 
encountered also in several two-dimensional quantum-gravity 
models.\cite{AN2,AN3,AN4,AN5}
\par
The BRS Noether current \eqno(2,18) can be rewritten as
\begin{eqnarray}
&&j_b{}^\mp=j_b'{}^\mp + (\tb^\pm-\tcalT^\pm)c^\pm, \\
\noalign{\noindent with}
&&j_b'{}^\mp \equiv -\tb^\pm c^\pm + i\barc^\pm c^\pm \partial_\pm c^\pm.
\end{eqnarray}
At the operator level, $j_b'{}^\mp$ strictly equals $j_b{}^\mp$.
But this equality no longer holds at the representation level because of
the appearance of the field-equation anomaly.  Indeed, $j_b{}^\mp$ is 
anomalous for $D\not=26$, while $j_b'{}^\mp$ is free of anomaly for 
any value of $D$.\cite{AN4,AN5}  On the other hand, $j_c{}^\mp$ is free 
of anomaly without making any modification.

\section{Perturbative approach}
The perturbative approach is so familiar to everybody that no explanation 
about it is necessary.  Nevertheless, when compared with the 
operator-formalism approach, the perturbative approach is seen to yield 
some surprising results.
\par
The Lagrangian density $\lag$ is decomposed into the free one $\lag_0$, 
which is quadratic with respect to the fields adopted as the basic ones,
and the remainder $\lagI$, called the interaction Lagrangian density.
\par
The Feynman propagators are obtained by taking the inverse of the 
differential operator sandwiched by the fields in $\lag_0$.
Thus, from \eqno(2,13), we have the following nonvanishing Feynman 
propagators:\foot{The subscript 0 indicates that the propagators are
free ones.}
\begin{eqnarray}
&&\taustar{\phi_M(x_1)\phi^N(x_2)}_0=\delta_M{}^N\Df(x_1-x_2), 
     \label{eq:4,1}\\
&&\taustar{\tb^\pm(x_1)h_\pm(x_2)}_0=-2i\delta^2(x_1-x_2),     
     \label{eq:4,2}\\
&&\taustar{\barc^\pm(x_1)c^\pm(x_2)}_0=-2i\partial_\pm\Df(x_1-x_2), 
  \label{eq:4,3}
\end{eqnarray}
where
\begin{eqnarray}
\partial_\pm\Df(x)&\equiv&\theta(x^0)\partial_\pm\Dp(x)
                         +\theta(-x^0)\partial_\pm[\Dp(x)]^* 
 \nonumber\\
&=&-{1\over 4\pi}\bigg[ {\theta(x^\mp+x^\pm)\over x^\pm-i0}
                       +{\theta(-x^\mp-x^\pm)\over x^\pm+i0} \bigg]. 
 \label{eq:4,4}
\end{eqnarray}
\par
It is quite remarkable that $\taustar{\tb^\pm\,h_\pm}_0$ is 
{\it nonvanishing\/} in spite of the fact that $h_\pm$ is a {\it zero\/}
operator as is seen from \eqno(3,1).  In contrast with the Wightman 
functions, the T$^*$-product does not respect the validity of the
field equations.  As is seen from \eqno(4,3), it is also inadmissible to 
set $\partial_\mp c^\pm=\partial_\mp\barc^\pm=0$ in the perturbative 
approach.  Hence we cannot discard the terms involving 
$\partial_\mp\barc^\pm$  in $\lagI$, that is, we have to distinguish
$\hcalT^\pm$ from $\tcalT^\pm$.  Thus the beautiful result of the operator 
formalism that $c^\pm$, $\barc^\pm$, $\tb^\pm$ and $\partial_\pm\phi_M$
are irrelevant to $x^\mp$ is no longer valid in the perturbative approach.
This fact makes the perturbative calculation complicated and sometimes
misleading, as we shall see later.
\par
By using $\lagI$ given by \eqno(2,14), we can easily calculate the 
$n$-point Green's functions.  For example, we have
\begin{eqnarray}
&&\taustar{\phi_M(x_1)\tb^\pm(x_2)\phi^N(x_3)}
=-2\delta_M{}^N\partial_\pm\Df(x_1-x_2)\cdot\partial_\pm\Df(x_2-x_3), 
 \label{eq:4,5} \\
&&\taustar{c^\pm(x_1)\tb^\pm(x_2)\barc^\pm(x_3)}
=8i\partial_\pm\Df(x_1-x_2)\cdot\partial_\pm{}^2\Df(x_2-x_3) 
    \nonumber \\
&&\hspace*{135pt}
 -4i\partial_\pm{}^2\Df(x_1-x_2)\cdot\partial_\pm\Df(x_2-x_3),  
 \label{eq:4,6}\\
&&\taustar{c^\pm(x_1)\tb^\pm(x_2)\barc^\mp(x_3)}
=-2\delta^2(x_1-x_2)\partial_\mp\Df(x_2-x_3). \label{eq:4,7}
\end{eqnarray}
Evidently, \eqno(4,5) and \eqno(4,6) correspond to \eqno(3,14) and to
\eqno(3,15), respectively.  However, \eqno(4,7) is a result peculiar to
the T$^*$-product.  This result is seen to be consistent with the 
Ward-Takahashi identity
\begin{eqnarray}
&&\taustar{\gdel(c^\pm(x_1)\barc^\pm(x_2)\barc^\mp(x_3))}=0,
\end{eqnarray}
because the second term of 
$\gdel c^\pm=-c^\pm\partial_\pm c^\pm-c^\mp\partial_\mp c^\pm$
contributes.
\par
Now, we come to the crucial point.  In sharp contrast with the case of 
the operator-formalism approach, the perturbative approach yields quite 
a nontrivial result for the $n$-point Green's function consisting of 
B-fields only.  Indeed, its connected part is given by a sum over 
one-loop Feynman diagrams.  
For example, we consider $\taustar{\tb^\lambda(x_1)\tb^\rho(x_2)}$.
Because of the nonvanishing of \eqno(4,2), the second-order perturbation
term yields
\begin{eqnarray}
&&\taustar{\tb^\lambda(x_1)\tb^\rho(x_2)}
=\taustar{\hcalT^\lambda(x_1)\hcalT^\rho(x_2)}_0.  \label{eq:4,9}
\end{eqnarray}
Therefore, we have
\begin{eqnarray}
&&\taustar{\tb^\pm(x_1)\tb^\pm(x_2)}
 =2(D-26)[\partial_\pm{}^2\Df(x_1-x_2)]^2,
 \label{eq:4,10} \\
&&\taustar{\tb^\pm(x_1)\tb^\mp(x_2)}=-{D-2\over2}[\delta^2(x_1-x_2)]^2.
 \label{eq:4,11}
\end{eqnarray}
They are divergent and therefore require the introduction of counter terms.
Note that the use of the T$^*$-product is responsible for the appearance 
of these divergences.
\par
The nonvanishing of the Green's functions consisting of B-fields only 
implies the violation of the BRS invariance.  
In the de Donder gauge case, Takahashi\cite{Takahashi} proposed to 
convert the violation of the BRS invariance for $D\not=26$ in the 
two-point B-field Green's function into the conformal 
anomaly.\foot{He made no mention 
about how to remove the BRS violation in the {\it higher-point\/} functions.}
We apply his line of thought to the present model.  
In addition to \eqno(4,10) and \eqno(4,11), we must take it into account 
the following exact two-point Green's functions:
\begin{eqnarray}
&&\taustar{h_\lambda(x_1)h_\rho(x_2)}=0, \\
&&\taustar{\tb^\lambda(x_1)h_\rho(x_2)}
 =-2i\delta_\rho{}^\lambda\delta^2(x_1-x_2). \label{eq:4,13}
\end{eqnarray}
The two-point functions of the effective action $\Gamma$ is obtained by 
taking the matrix inverse of \eqno(4,10)-\eqno(4,13).  
Accordingly, we have
\begin{eqnarray}
\Gamma&=&\int d^2x_1\int d^2x_2\bigg[
 -{1\over\,2\,}\delta^2(x_1-x_2)\tb^\lambda(x_1)h_\lambda(x_2) \nonumber\\
&& \qquad 
+{D-26\over2}i\sum_{\alpha=\pm}
 ([\partial_\alpha{}^2\Df(x_1-x_2)]^2)_{\hbox{\tiny R}}
 h_\alpha(x_1)h_\alpha(x_2)
+ \ \cdots\ \bigg], \label{eq:4,14} 
 \ifpreprint  \else \hspace*{30pt}\fi
\end{eqnarray}
where a subscript R indicates regularization.  The BRS-violating term 
in \eqno(4,14) is converted into the conformal-anomaly term by adding 
the conformal degree of freedom.  We do not work out this procedure 
in detail because it is not our aim to do so.
\par
The important point is the violation of the BRS invariance in the B-field
Green's functions.  In the de Donder gauge case, the BRS violation has 
arisen by applying the dimensional regularization only to internal lines
but {\it not to external lines}.\cite{AN6,AN8}
In the present model, external lines are absent because \eqno(4,2) is 
local. Instead, as is seen from \eqno(4,9), perturbative approach makes 
use of the field-equation anomaly without being aware of this fact. 

\section{Path-integral approach}
The path integral directly gives us the generating function of the Green's
functions, that is, it deals with the T$^*$-product quantities only.
The path integral $Z$ is formally expressed as
\begin{eqnarray}
&&Z(J)=\int \big(\prod_i{\cal D}\vphi_i\big)\,
                  \exp\, i\!\int d^N\!{}x(\lag+\sum_iJ_i\vphi_i)
\end{eqnarray}
with $Z(0)=1$, where $J_i$ denotes the source function corresponding to 
the field $\vphi_i$.
\par
It is possible to derive the path-integral formula [corrected by 
the Lee-Yang term proportional to $\delta^N(0)$] from the canonical 
operator formalism.\cite{Buchbinder}
In this sense, the path-integral formalism can be regarded as the one
equivalent to the operator formalism.  But, one should note that, 
in this derivation, one must use the {\it field equations at the 
representation level}.  
This fact implies that the path-integral formalism cannot take care 
of the existence of the field-equation anomaly.
\par
From the successful experience of discussing the anomaly problem in 
gauge theories, it has been customary to believe that any anomaly always
arises from the non-invariance of the path-integral measure under 
the symmetry which leaves the action $S$ invariant.  
But we point out that anomalies can arise also from the field-equation 
anomaly which is beyond the scope of the path-integral formalism.
\par
Let $F(\vphi)$ be an arbitrary function of $\vphi_i$'s.  
The path-integral measure is supposed to be invariant under the functional 
translation $\vphi_i \longrightarrow \vphi_i +\delta\vphi_i$.
Hence, by considering a variation of a field $\vphi_i$ in 
$F(i^{-1}\partial/\partial J)Z|_{J=0}$, we obtain
\begin{eqnarray}
&&i\taustar{F(\vphi){\delta S\over \delta \vphi_i}}
+\taustar{{\delta F(\vphi)\over \delta \vphi_i}} =0. \label{eq:5,2}
\end{eqnarray}
This equation corresponds to the field equation 
$\delta S/\delta \vphi_i=0$ of the operator formalism.  
The second term of \eqno(5,2) is a field-equation
violating term due to the use of the T$^*$-product.  
One should never confuse it with the field-equation anomaly.  
For example, in the conformal-gauge two-dimensional quantum gravity, 
\eqno(4,14) is reproduced from \eqno(5,2) by setting $F=\tb^\lambda$ 
and $\vphi_i=\tb^\rho$.  Likewise, if we set $F=\tb^\lambda$ and 
$\vphi_i=h_\rho$ in \eqno(5,2), we obtain
\begin{eqnarray}
&&\taustar{\tb^\lambda(\tb^\rho-\hcalT^\rho)}=0, \label{eq:5,3}
\end{eqnarray}
that is, we do not encounter the field-equation anomaly.
Instead, as is shown in the perturbative approach, \eqno(5,3) induces 
the violation of the BRS invariance in the path-integral approach.
\par
Historically, the anomaly problem in the conformal-gauge two-dimensional
quantum gravity was discussed first by Fujikawa\cite{Fujikawa} in the 
path-integral formalism.  
His formulation is not, however, understandable in the framework stated
in \S4.
\par
First, he takes all {\it three\/} degrees of freedom of $g_{\mu\nu}$ 
as path-integration variables; nevertheless, each of his ghost fields 
has only {\it two\/} degrees of freedom. The extra one degree of freedom
is the conformal one, denoted by $\rho$, is 
{\it not allowed\/}\foot{If one carries out the integration over 
$\rho$ after introducing the
conformal ghosts, then it becomes impossible to work out his analysis.}
to be integrated (until the Liouville action is derived) in spite of the
fact that it is an independent path-integration variable.
\par
Second, by introducing tilde fields $\tilde\vphi{}_i=\rho^{n_i}\vphi_i$,
$n_i$ being a certain fractional number, he claims that the path-integral 
measure becomes BRS invariant if it is expressed in terms of the tilde 
fields.  He then derives the Liouville action expressed in terms of 
$\rho$ alone by calculating the variation of the path-integral measure 
under the conformal transformation. That is, according to his theory, 
the conformal anomaly is directly obtained {\it without passing 
through the BRS anomaly\/} in contradiction to the consideration
presented in \S4.  
\par
We are thus unable to reproduce his analysis in terms of the explicit
solution.

\section{FP-ghost number current anomaly}
As we emphasized previously,\cite{AN4} there is no FP-ghost number anomaly
in the conformal-gauge two-dimensional quantum gravity: The exact solution
is completely consistent with the FP-ghost number conservation.
The conservation of the FP-ghost number current $j_c{}^\mu$ is a simple
consequence of the fact that $c^\pm(x)$ and $\barc^\pm(x)$ are independent
of $x^\mp$.  This property is never violated at the representation level.
Nevertheless, many authors have claimed that the FP-ghost number current
has anomaly.  The reasons for the occurrence of this belief are its
correspondence to the Riemann-Roch theorem and the field-equation-violating
property of the T$^*$-product.
\par
Fujikawa\cite{Fujikawa} was the first to claim the existence of the 
FP-ghost number current anomaly.  He derived it by making the FP-ghost 
number transformation\foot{This must be done in the original 
tilde-variable expression.  No anomaly is derived by the naive calculation
if this transformation is made in the equivalent expression having the 
Liouville action explicitly.} in his path-integral formalism described
at the end of \S5.  His result is written as
\begin{eqnarray}
&&\partial_\lambda\langle\!\langle\,j_c{}^\lambda\,\rangle\!\rangle
 ={3\over 4\pi}\langle\!\langle\, \partial^2 \log\rho \,\rangle\!\rangle,
 \label{eq:6,1}
\end{eqnarray}
where we denote the path integration by 
$\langle\!\langle\;\cdots\;\rangle\!\rangle$.
If the degrees of the reparametrization freedom is suppressed, one may 
write $-\partial^2\log\rho=\sqrt{g}R$ (Euclidean metric is used).
Here we must note that Fujikawa's theory is formulated in the {\it flat\/}
background metric and that $\rho$ is the path-integration variable.
\par
Shortly later, Friedan, Martinec and Shenker,\cite{FMS} who formulated
conformal field theory, quoted \eqno(6,1) in the disguised form.
They consider a completely {\it curved\/} background metric $\hg_{\mu\nu}$.
The right-hand side of their equation is const.$\sqrt{\hg}\hat{R}$, 
{\it a function of\/} $\hg_{\mu\nu}$, which is nothing but the quantity 
required by the Riemann-Roch theorem under the prerequisite of the 
conformal covariance.
It may be certainly {\it analogous\/} to \eqno(6,1), but we cannot find any
{\it logical connection\/} between them.\foot{According to Fujikawa
(private communication), the possible existence of the zero points of $\rho$
can take care of the effect of the topological number $\int d^2x\,\sqrt{\hg}
\hat{R}$.  We do not see how this idea can be formulated.}
\par
In the perturbative approach, the Friedan-Martinec-Shenker version of 
\eqno(6,1) is interpreted, through the consideration based on the effective
action, as the matter that\cite{Dusedau,KR}
\begin{eqnarray}
&&J^\lambda{}_{\mu\nu}\equiv\taustar{j_c{}^\lambda 
  {\delta S\over \delta \hg^{\mu\nu}}}|_{\hg^{\mu\nu}=\eta^{\mu\nu}}
  \label{eq:6,2}
\end{eqnarray}
has a nonvanishing nonlocal term, where $\hg_{\mu\nu}$ is a background
metric introduced in such a way that the gauge-fixing plus FP-ghost 
Lagrangian density becomes background covariant.
Note in \eqno(6,2) that the background metric is taken to be {\it flat\/}
in the Feynman-diagram calculation.
\par
In the conformal-gauge case, $J^\lambda{}_{\mu\nu}$ is essentially equal to
\begin{eqnarray}
&&\taustar{j_c{}^\pm(x_1)\tcalT^\pm(x_2)}
 =-12\partial_\pm\Df(x_1-x_2)\cdot\partial_\pm{}^2\Df(x_1-x_2). 
 \label{eq:6,3}
\end{eqnarray}
It is in this sense that the FP-ghost number current anomaly is claimed 
to be obtained in the perturbative approach.
It should be noted, however, that {\it if the T$^*$-product is not taken,
that is, if $\Df$ is replaced by $\Dp$, \eqno(6,3) becomes consistent with\/}
$\partial_\lambda j_c{}^\lambda=0$.
\par
D\"usedau\cite{Dusedau} found that $J^\lambda{}_{\mu\nu}$ has no nonlocal
term in the de Donder-gauge case.  Kraemmer and Rebhan\cite{KR} discussed
the gauge dependence of $J^\lambda{}_{\mu\nu}$ and claimed that the gauge
independence can be recovered if one adds a contribution from the
``Lagrange-multiplier (or B-field) current''.\foot{Although we cannot
regard their proof as adequate, their claim itself can be verified by 
explicit calculation.\cite{AN9}}  But the relation between this fact and 
the Riemann-Roch theorem was not discussed.
\par
Recently, Takahashi\cite{Takahashi} has reconsidered D\"usedau's analysis
from his perturbative approach described in \S4.  
In discussing $j_c{}^\lambda$, he regards the quantum gravitational field
as the background metric, just as Fujikawa did.  Rederiving D\"usedau's
result, he asserts that the vanishing of the FP-ghost number current 
anomaly can be explained by the existence of the FP-ghost conjugation
invariance of the de Donder gauge two-dimensional quantum gravity.
One should note, however, that, as pointed out in \S2, the {\it FP-ghost
conjugation invariance exists also in the conformal-gauge case}.
Therefore, his standpoint would imply the absence of the FP-ghost number
current anomaly also in the conformal-gauge case.
\par
Finally, we note that the nonexistence of the FP-ghost number current 
anomaly can be shown even if the gauge-fixing background metric is 
a nonflat one given by $\hat\rho(x)\eta_{\mu\nu}$, where 
$\hat\rho{}^{-1}$ {\it is assumed to exist}. 
In this case, \eqno(2,6) is replaced by
\begin{eqnarray}
&&\tg^{\mu\nu}=(\eta^{\mu\nu}+\hat\rho^{-1} h^{\mu\nu})
               (1-\hat\rho{}^{-2}\det{h^{\sigma\tau}})^{-1/2}.
\end{eqnarray}
The Lagrangian density of this case is obtained from \eqno(2,8) by
simply replacing $h_\lambda$ by $\hat\rho^{-1}h_\lambda$.
Since $\gdel(\hat\rho^{-1}h_\lambda)=\hat\rho^{-1}\gdel(h_\lambda)$,
we can absorb the factor $\hat\rho^{-1}$ into $\tb^\lambda$ and 
$\barc^\lambda$ by redefining them. Thus the ghost part of 
the Lagrangian density of the nonflat case becomes completely the same
as that of the flat case.  Thus nothing new can happen about the 
FP-ghost number current.

\section{Discussion}
Nowadays, the path-integral approach and the counter-term business have
become so fashionable that many physicists preclude the consideration 
based on other approaches from the outset.  Certainly, the path-integral 
approach is convenient and successful in gauge theories, but we wish to 
emphasize that the same is not necessarily true in quantum gravity.
\par
The path-integral formalism directly deals with the solution at the 
representation level.  Accordingly, if one adopts the path-integral 
approach, one can no longer perceive what happens in the transition from
the operator level to the representation level.  Indeed, one cannot
describe the existence of the field-equation anomaly in the path-integral
approach.
\par
The quantities describable by the path-integral formalism are those
which can be written in terms of the T$^*$-product.  The T$^*$-product is
certainly a very convenient notion because we need not take care of the 
ordering problem even for the timelikely separated field operators.
On the other hand, as emphasized in the present paper, the T$^*$-product
has a demerit of violating the field equations explicitly.
As demonstrated in the present paper, this fact induces unpleasant 
complications and misleading expressions.
Furthermore, since the T$^*$-product contains $\theta$-functions, the 
Green's function is more singular than the corresponding Wightman functions,
that is, some singularities found in the perturbative or path-integral
approach may be superficial.  When this fact is combined with the 
counter-term business, one is led to introducing counter terms which 
are purely of the T$^*$-product origin. In the present paper, we have 
demonstrated that ``anomalies'' also can be of the T$^*$-product origin.
\par
We hope that more physicists reinvestigate the anomaly problem in 
quantum gravity without adhering to the path-integral approach.

\section*{Acknowledgements}
One of the present authors (N. N.) would like to express his sincere
thanks to Professor K. Fujikawa and Dr. H. Kanno for the discussions 
concerning the FP-ghost number current anomaly.

%
%
 
\ifpreprint
\newpage
\fi

\end{document}